\documentclass[aps,prb,showpacs]{revtex4-1}

\usepackage{graphicx}

\begin{document}
\narrowtext
\title{Smooth electron waveguides in graphene}

\author{R. R. Hartmann}
\affiliation{School of Physics, University of Exeter, Stocker Road,
Exeter EX4 4QL, United Kingdom}

\author{N. J. Robinson}
\affiliation{School of Physics, University of Exeter, Stocker Road,
Exeter EX4 4QL, United Kingdom}

\author{M. E. Portnoi}
\email[]{m.e.portnoi@exeter.ac.uk} \affiliation{School of Physics,
University of Exeter, Stocker Road, Exeter EX4 4QL, United Kingdom}

\date{21 June 2010}

\begin{abstract}
We present exact analytical solutions for the zero-energy modes of
two-dimensional massless Dirac fermions fully confined within a
smooth one-dimensional potential $V(x)=-\alpha/\cosh(\beta{}x)$,
which provides a good fit for potential profiles of existing
top-gated graphene structures. We show that there is a threshold
value of the characteristic potential strength $\alpha/\beta$ for
which the first mode appears, in striking contrast to the
non-relativistic case. A simple relationship between the
characteristic strength and the number of modes within the potential
is found. An experimental setup is proposed for the observation of
these modes. The proposed geometry could be utilized in future
graphene-based devices with high on/off current ratios.
\end{abstract}

\pacs{73.21.-b, 03.65.Ge, 03.65.Pm, 81.05.Uw}

\maketitle

\section{Introduction}
Klein proposed that relativistic particles do not experience
exponential damping within a barrier like their non-relativistic
counterparts, and that as the barrier height tends towards infinity,
the transmission coefficient approaches unity.\cite{Klein}
This inherent property of relativistic particles makes confinement
non-trivial. Carriers within graphene behave as two-dimensional (2D)
massless Dirac fermions, exhibiting relativistic behavior at
sub-light speed\cite{DiracFermions,CastroNetoReview} owing to their
linear dispersion, which leads to many optical analogies.
\cite{Lens,LevitovParabolic,ZBZcombined,Beenakker_PRL_102_2009,Chen_APL_94_2009}
Klein tunneling through
p-n junction structures in graphene has been studied both theoretically
\cite{LevitovParabolic,KleinCombined,Cheianov_Falko_PRB(R)_74_2006,Peeters_PRB_74_2006,Chaplik_JETP_84_2006,Peeters_APL_90_2007,Fogler_PRL_100_2008,Fogler_PRB_77_2008,BeenakkerRMP08,ChineseShape} and experimentally.
\cite{Transport PN,TopGateCombined,Kim_PRL_99_2007,Savchenko_NanoLett_8_2008,Liu_APL_92_2008,GG_PRL_102_2009,Kim_NatPhys_5_2009} Quasi-bound states were considered in order to study resonant tunneling through various
sharply terminated barriers.
\cite{LevitovParabolic,Peeters_PRB_74_2006,Chaplik_JETP_84_2006,Peeters_APL_90_2007,ChineseShape}
We propose to change the geometry of the problem in order to study the
propagation of fully confined modes along a smooth electrostatic
potential, much like photons moving along an optical fiber.

\begin{figure}[h]
  \centering
    \includegraphics[width=7.5cm]{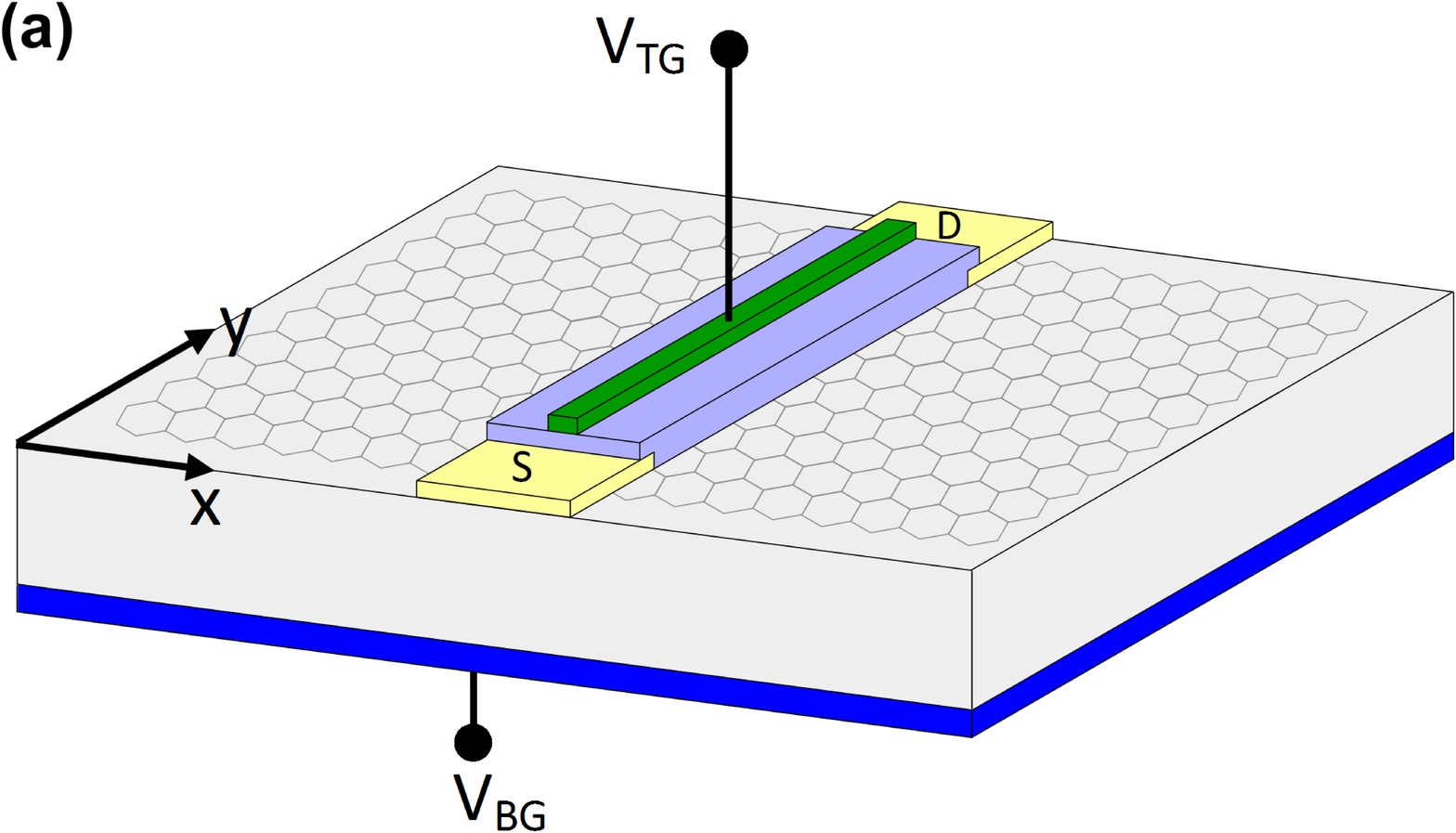}
    \includegraphics[width=7.5cm]{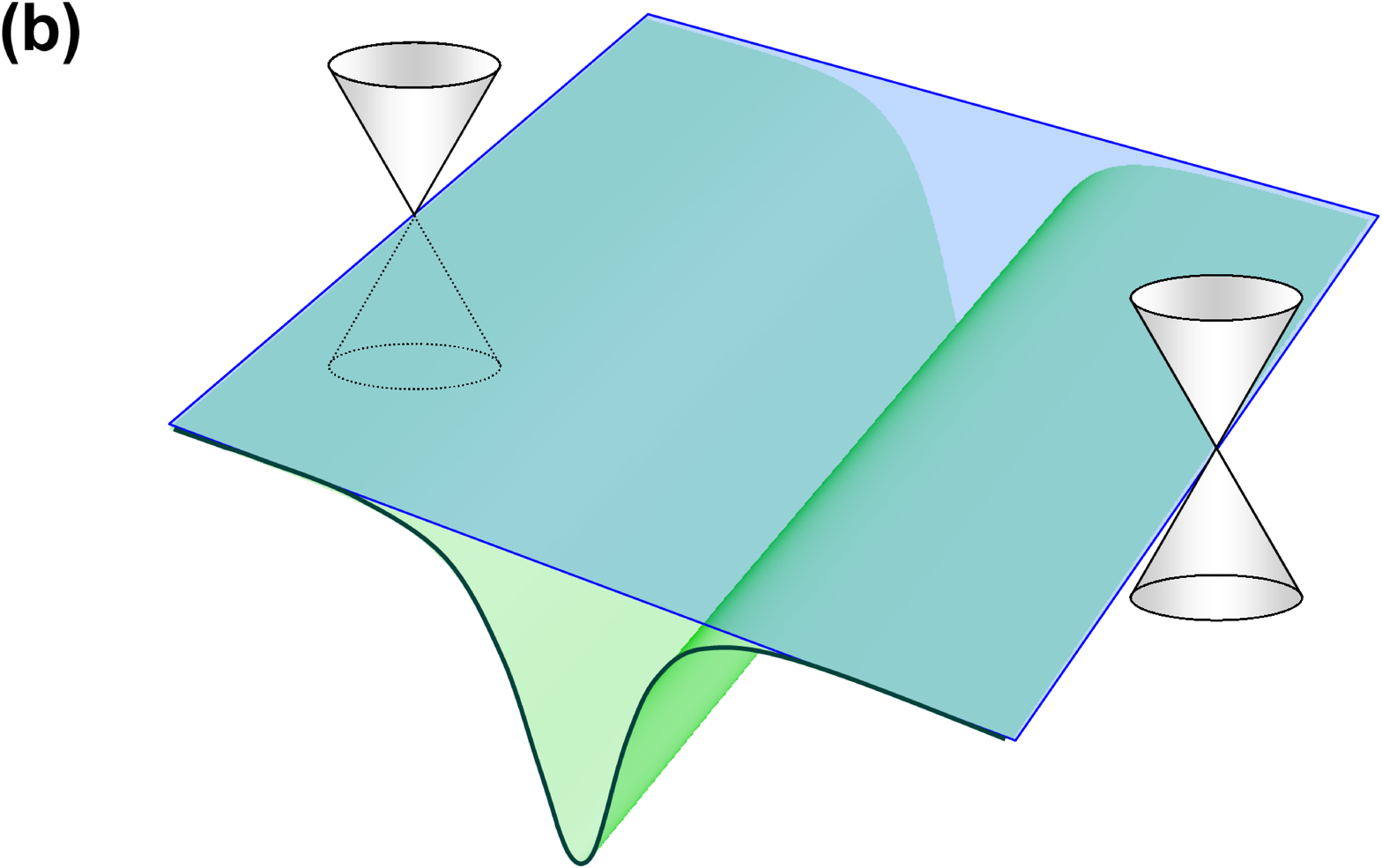}
      \caption{(a) A schematic diagram of a Gedankenexperiment
        for the observation of localized modes in graphene waveguides,
        created by the top gate (V$_{\textrm{\scriptsize{TG}}}$). The Fermi level is set using the back gate
        (V$_{\textrm{\scriptsize{BG}}}$) to be at the Dirac point ($\varepsilon_{\textrm{\scriptsize{F}}}=0$).
        (b) The electrostatic potential created by the applied top gate voltage.
        The plane shows the Fermi level position at $\varepsilon_{\textrm{\scriptsize{F}}}=0$.}
  \label{fig:Cosh_1/2}
\end{figure}

So far quasi-one-dimensional channels have been achieved within
graphene nanoribbons,
\cite{CastroNetoReview,Nanoribbons,RibbonTheoryCombined,Efetov_PRL_98_2007,Peres_JPhysCondMat_21_2009}
however, controlling their transport properties requires precise tailoring
of edge termination,\cite{RibbonTheoryCombined} currently
unachievable. In this paper we claim that truly bound modes can be
created within bulk graphene by top gated structures,
\cite{Kim_PRL_99_2007,Savchenko_NanoLett_8_2008,Liu_APL_92_2008,GG_PRL_102_2009,Kim_NatPhys_5_2009} such as the one shown in
Fig.~\ref{fig:Cosh_1/2}(a). In an ideal graphene sheet at
half-filling, the Fermi level is at the Dirac point and the density
of states for a linear 2D dispersion vanishes. In realistic graphene
devices the Fermi level can be set using the back gate. This is key
to the realization of truly bound modes within a graphene waveguide,
as zero-energy modes cannot escape into the bulk as there are no
states to tunnel into. Moreover the electrostatic confinement
isolates carriers from the sample edges, which are considered as a
major source of intervalley scattering in clean graphene.\cite{SavchenkoSSC09}

In this paper we obtain an exact analytical solution for bound
modes within a smooth electrostatic potential in pristine graphene at half-filling, count
the number of modes and calculate the conductance of the channel.
The conductance carried by each of these modes is comparable to the
minimal conductivity of a realistic disordered graphene system.
\cite{DiracFermions,Min. Con1,MinConTheoryCombined,DasSarma_PNASUSA_104_2007}
For the considered model potential we show that there is a threshold
potential characteristic strength (the product of the potential
strength with its width), for which bound modes appear. Whereas a
symmetric quantum well always contains a bound mode for
non-relativistic particles, we show that it is not the case for
charge carriers in graphene.

\section{Fully confined modes in a model potential}
The Hamiltonian of graphene for a two-component Dirac wavefunction
in the presence of a one-dimensional potential $U(x)$ is
\begin{equation}
\hat{H}=v_{\textrm{\scriptsize{F}}}\left(\sigma_{x}\hat{p}_{x}+\sigma_{y}\hat{p}_{y}\right)+U(x),
\label{eq:Hamiltonian}
\end{equation}
where $\sigma_{x,y}$ are the Pauli spin matrices,
$\hat{p}_{x}=-i\hbar\frac{\partial}{\partial x}$ and
$\hat{p}_{y}=-i\hbar\frac{\partial}{\partial y}$ are the momentum
operators in the $x$ and $y$ directions respectively and
$v_{\textrm{\scriptsize{F}}}\approx1\times10^{6}$m/s is the Fermi velocity in graphene.
In what follows we will consider smooth confining potentials,
which do not mix the two non-equivalent valleys.
All our results herein can be easily reproduced for the other valley.
When Eq.~(\ref{eq:Hamiltonian}) is applied to a
two-component Dirac wavefunction of the form:
\[
\mbox{e}^{iq_{y}y}\left({\Psi_{A}(x) \atop \Psi_{B}(x)}\right),
\]
where $\Psi_{A}(x)$ and $\Psi_{B}(x)$ are the wavefunctions associated
with the $A$ and $B$ sublattices of graphene respectively and the free
motion in the $y$-direction is characterized by the wavevector
$q_{y}$ measured with respect to the Dirac point, the following coupled
first-order differential equations are obtained:
\begin{equation}
\left(V(x)-\varepsilon\right)\Psi_{A}(x)-i\left(\frac{\mbox{d}}{\mbox{d}x}+q_{y}\right)\Psi_{B}(x)=0,\label{eq:basic1}
\end{equation}
\begin{equation}
-i\left(\frac{\mbox{d}}{\mbox{d}x}-q_{y}\right)\Psi_{A}(x)+\left(V(x)-\varepsilon\right)\Psi_{B}(x)=0.\label{eq:basic2}
\end{equation}
Here $V(x)=U(x)/\hbar v_{\textrm{\scriptsize{F}}}$ and energy $\varepsilon$ is measured in
units of $\hbar v_{\textrm{\scriptsize{F}}}$.

For the treatment of confined modes within a symmetric electron
waveguide, $V(x)=V(-x)$, it is convenient to consider symmetric and
anti-symmetric modes. One can see from
Eqs.~(\ref{eq:basic1}-\ref{eq:basic2}) that $\Psi_{A}(x)$ and
$\Psi_{B}(x)$ are neither even nor odd, so we transform to
symmetrized functions:
\[
\Psi_{1}=\Psi_{A}(x)-i\Psi_{B}(x),\quad\ensuremath{\Psi_{2}}=\ensuremath{\Psi_{A}(x)+i\Psi_{B}(x)}.\]
The wavefunctions $\Psi_{1}$ and $\Psi_{2}$ satisfy the following
system of coupled first-order differential
equations:\begin{equation}
\left[V(x)-\left(\varepsilon-q_{y}\right)\right]\Psi_{1}-\frac{\mbox{d}\Psi_{2}}{\mbox{d}x}=0,\label{eq:sym1}\end{equation}
\begin{equation}
\frac{\mbox{d\ensuremath{\Psi_{1}}}}{\mbox{d}x}+\left[V(x)-\left(\varepsilon+q_{y}\right)\right]\Psi_{2}=0.
\label{eq:sym2}
\end{equation}
It is clear from Eqs.~(\ref{eq:sym1}-\ref{eq:sym2}) that $\Psi_1$
and $\Psi_2$ have opposite parity.

For an ideal graphene sheet at half-filling the conductivity is
expected to vanish due to the vanishing density of states. When the
Fermi energy is at the Dirac point ($\varepsilon=0$) there are no
charge carriers within the system, so graphene is a perfect
insulator. However all available experiments demonstrate
non-vanishing minimal conductivity
\cite{DiracFermions,CastroNetoReview,Min. Con1} of the order of
$e^{2}/{h}$ which is thought to be due to disorder within the system
\cite{CastroNetoReview,MinConTheoryCombined,DasSarma_PNASUSA_104_2007,PercolationNetwork} or
finite-size effects.\cite{SizeEffectCombined,Beenakker_PRL_96_2006}
In order to study confined states within and conductance along an
electron waveguide it is necessary to use the back gate to fix the
Fermi energy ($\varepsilon_{\textrm{\scriptsize{F}}}$) at zero, as shown in
Fig.~\ref{fig:Cosh_1/2}(b).
Note that Fig.~\ref{fig:Cosh_1/2}(a) is just a schematic
of the proposed experimental geometry
and that side contacts may be needed to maintain the
`bulk' Fermi level at zero energy.
The conductivity of the graphene sheet
is a minimum and for a square sample the conductance
is of the order of the conductance carried by a
single mode within a waveguide. Thus the appearance of
confined modes within the electron waveguide will drastically change
the conductance of a graphene flake. Indeed each mode will
contribute $4e^{2}/{h}$, taking into account valley and spin degeneracy.
For device applications the sample should be designed in such a way
that the contribution to the conductance from the confined modes is most prominent.
In the ideal case, the conductivity of the channel would
be the only contribution to that of the graphene sheet. When
$\varepsilon_{\textrm{\scriptsize{F}}}\neq0$, the conductivity will be dominated by the 2D
Fermi sea of electrons throughout the graphene sheet. Henceforth,
we shall consider the modes for $\varepsilon=0$.

We shall consider truly smooth potentials, allowing us to avoid the
statement of ``sharp but smooth'' potentials, which is commonly used
to neglect intervalley mixing for the tunneling problem.
\cite{KleinCombined,Cheianov_Falko_PRB(R)_74_2006,Peeters_PRB_74_2006,Chaplik_JETP_84_2006,Peeters_APL_90_2007}
Furthermore, we are interested in potentials that vanish at infinity and have at least
two fitting parameters, characterizing their width and strength, in
order to fit experimental potential profiles.
\cite{TopGateCombined,Kim_PRL_99_2007,Savchenko_NanoLett_8_2008,Liu_APL_92_2008,GG_PRL_102_2009,Kim_NatPhys_5_2009}
Let us consider the following class of potentials, which satisfy the aforementioned
requirements:\begin{equation}
V(x)=-\frac{\alpha}{\cosh^{\lambda}(\beta
x)},\label{eq:potential}\end{equation} where $\alpha$, $\beta$ and
$\lambda$ are positive parameters. The negative sign in
Eq.~(\ref{eq:potential}) reflects a potential well for electrons,
and similar results can easily be obtained for holes by changing the
sign of $V(x)$. Notably $\lambda=2$ is the familiar case of the
P\"{o}schl-Teller potential, which has an analytic solution for the
non-relativistic case.\cite{PoschlTeller}
It is shown in the Appendix that the model potential (\ref{eq:potential})
with $\lambda=1$ provides an excellent fit to a graphene top-gate structure.

Eliminating $\Psi_{2}$ ($\Psi_{1}$) reduces the system of
Eqs.~(\ref{eq:sym1}-\ref{eq:sym2}) to a single second order
differential equation for $\Psi_{1}$ ($\Psi_{2}$), which for the
potential given by Eq.~(\ref{eq:potential}) and $\varepsilon=0$ becomes
\begin{equation}
\frac{\mbox{d}^{2}\Psi_{1,2}}{\mbox{d}z^{2}}+\lambda\tanh(z)
\frac{\omega\cosh^{-\lambda}(z)}{\omega\cosh^{-\lambda}(z) \pm
\Delta}\frac{\mbox{d\ensuremath{\Psi_{1,2}}}}{\mbox{d}z}
+\left(\omega^{2}\cosh^{-2\lambda}(z)-\Delta^{2}\right)
\Psi_{1,2}=0, \label{eq:FinalDiff}
\end{equation}
where we use the dimensionless variables $z=\beta x$, $\Delta= q_{y}/{\beta}$
and $\omega=\alpha/\beta$.
For $\lambda=1$, the change of variable $\xi=\tanh(z)$ allows Eq.~(\ref{eq:FinalDiff})
to be reduced to a set of hypergeometric equations yielding the
following non-normalized bound solutions for $\Delta>0$:
\begin{eqnarray}
  \Psi_{1,2} &=&
  \pm{}\left(1+\xi\right)^{p}\left(1-\xi\right)^{q}\:_{2}\mbox{F}_{1}\left(p+q-\omega\mbox{, }
  p+q+\omega\mbox{; }2p+\frac{1}{2};\frac{1+\xi}{2}\right)
  \nonumber
  \\
  &+&
  (-1)^{n}\left(1+\xi\right)^{q}\left(1-\xi\right)^{p}\,_{2}
  \mbox{F}_{1}\left(p+q-\omega,\, p+q+\omega;\,2p+\frac{1}{2};\frac{1-\xi}{2}\right),
\label{eq:wfn1}
\end{eqnarray}
where in order to terminate the hypergeometric series it is
necessary to satisfy
$p=\frac{1}{2}\left(\omega-n\right)+\frac{1}{4}$,
$q=\frac{1}{2}\left(\omega-n\right)-\frac{1}{4}$ and
$\Delta=\omega-n-\frac{1}{2}$, where $n$ is a positive integer.
Though we have assumed that $\omega$ is positive, one can see that
the structure of the solutions in Eq.~(\ref{eq:wfn1}) remains
unchanged with the change of sign of $\omega$, reflecting
electron-hole symmetry. In order to avoid a singularity at
$\xi=\pm1$ we require that both $p>0$ and $q>0$ and obtain the
condition that $\omega-n>\frac{1}{2}$. It should be noted that this
puts an upper limit on $n$, the order of termination of the
hypergeometric series. Notably the first mode occurs at $n=0$, thus
there is a lower threshold of $\omega>\frac{1}{2}$ for which bound
modes appear. Hence within graphene, quantum wells are very
different to the non-relativistic case; bound states are not present
for any symmetric potential, they are only present for significantly
strong or wide potentials, such that $\omega=\alpha/\beta>\frac{1}{2}$.

\begin{figure}[h]
  \centering
     \includegraphics[width=6.0cm]{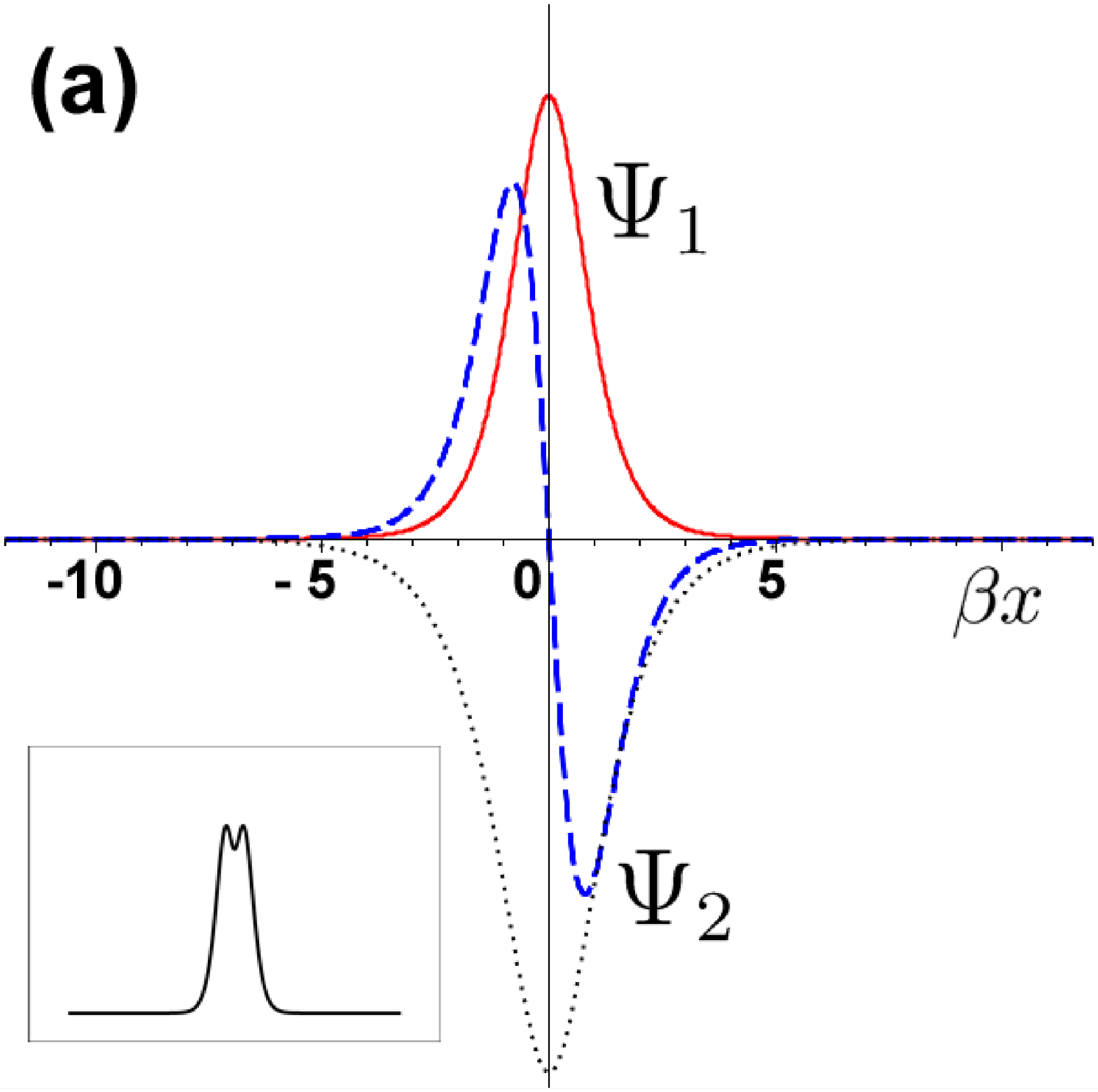}
    \includegraphics[width=6.0cm]{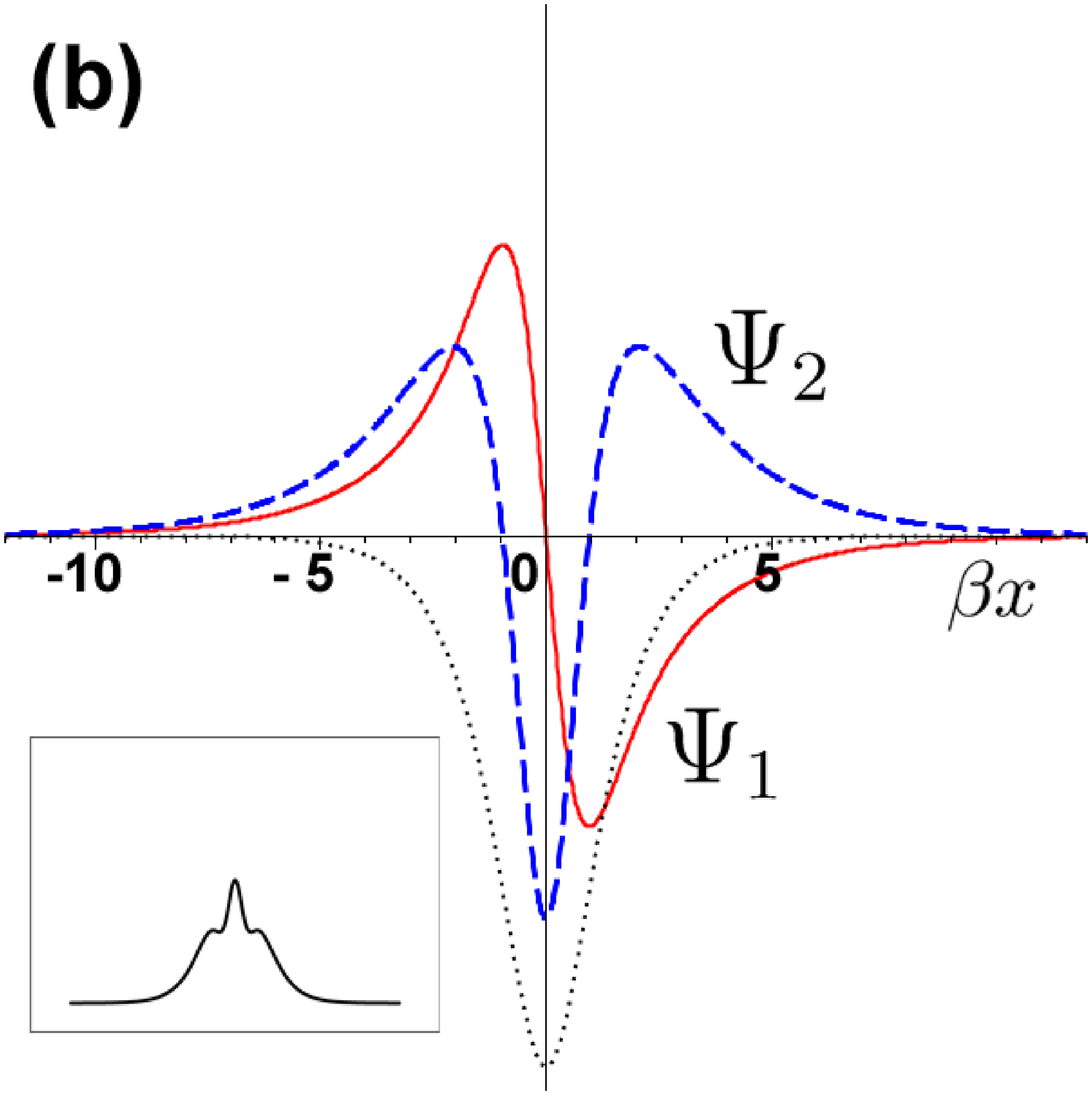}
        \caption{The wavefunctions $\Psi_1$ (solid line)
        and $\Psi_2$ (dashed line) are shown for $\omega=2$ for
        (a) the first ($n=0$) mode and (b) the second ($n=1$) mode.
        A potential profile is provided as a guide for the eye (dotted line).
        The insets show the electron density profile for the corresponding modes.}
  \label{fig:Cosh_3/4}
\end{figure}

Let us consider the first mode ($n=0$) in Eq.~(\ref{eq:wfn1}) which
appears within the electronic waveguide with increasing $\omega$. In
this case the hypergeometric function is unity, and the normalized
wavefunctions are:
\begin{equation}
\Psi_{1,2}= A_{1,2}\left[ \left(1+\xi\right)^{\frac{\omega}{2}-
\frac{1}{4}}\left(1-\xi\right){}^{\frac{\omega}{2}+\frac{1}{4}}\pm
\left(1+\xi\right)^{\frac{\omega}{2}+
\frac{1}{4}}\left(1-\xi\right){}^{\frac{\omega}{2}-
\frac{1}{4}}\right], \label{eq:wfnnorm}
\end{equation}
where $A_{1,2}$ is given by:
\begin{equation}
A_{1,2}=\left\{ \frac{\beta\,
\left(2\omega-1\right)\Gamma\left(\omega\right)\Gamma\left(\omega+\frac{1}{2}\right)}
{4\sqrt{\pi}\left[2\Gamma^{2}\left(\omega+\frac{1}{2}\right)
\pm\left(2\omega-1\right)\Gamma^{2}\left(\omega\right)\right]}
\right\}^{\frac{1}{2}},
\end{equation}
where $\Gamma(z)$ is the Gamma function. As expected, the two
functions given by Eq.~(\ref{eq:wfnnorm}) are of different parity,
thus unlike the non-relativistic case there is an odd function
corresponding to the first confined mode. This leads to a threshold
in the characteristic potential strength $\omega$ at which the first
confined mode appears; much like in the conventional quantum well,
where the first odd state appears only for a sufficiently deep or
wide potential well. In Fig.~\ref{fig:Cosh_3/4} we present
$\Psi_{1}$, $\Psi_{2}$ and the corresponding electron density
profiles for the first and second bound modes for the case of
$\omega=2$. The shape of the confinement potential is shown for
guidance within the same figure. The charge density profile for
these modes differs drastically from the non-relativistic case. The
first mode ($n=0$) has a dip in the middle of the potential well,
whereas the second mode ($n=1$) has a maximum. This is a consequence
of the complex two-component structure of the wavefunctions.

For negative values of  $\Delta$ , $\Psi_1$ and $\Psi_2$ switch
parity such that $\Psi_{1,2}(-\Delta)=\Psi_{2,1}(\Delta)$. This
means backscattering within a channel requires a change of parity of
the wavefunctions in the $x$-direction. Notably, when another
non-equivalent Dirac valley is considered, one finds that there are
modes of the same parity propagating in the opposite direction.
However intervalley scattering requires a very short-range potential
or proximity to the sample edges. Thus for smooth scattering
potentials backscattering should be strongly suppressed. Such
suppression should result in an increase in the mean free path of
the channel compared to that of graphene. This is similar to the
suppression of backscattering in carbon nanotubes,\cite{Ando} where
the ballistic regime is believed to persist up to room temperature
with a mean free path exceeding one micrometer.\cite{CNTCombined,Porntoi_NanoLett_7_2007}
In some sense the considered waveguide can be thought of as a carbon
nanotube-like structure with parameters controlled by the top gate.

Notably, our results are also applicable for the case of a
one-dimensional massive particle which is confined in the same
potential. This can be achieved by the substitution
$|q_y|\rightarrow m v_{\textrm{\scriptsize{F}}}/\hbar$ into
Eqs.~(\ref{eq:sym1}-\ref{eq:sym2}), where the gap is given by
$2m v_{\textrm{\scriptsize{F}}}^{2}$.
Hence in the massless 2D case the momentum along the
waveguide plays the same role as the gap in the massive
one-dimensional case. Therefore, in a massive one-dimensional Dirac
system (such as a narrow gap carbon nanotube) there exists a bound
state in the middle of the gap for certain values of the
characteristic strength of the potential.

The number of modes $N_{\omega}$ at a fixed value of $\omega$ is the
integer part of $\omega+\frac{1}{2}$ herein denoted
$N_{\omega}=\left\lfloor \omega+\frac{1}{2}\right\rfloor $. The
conductance of an ideal one-dimensional channel characterized by
$\omega$ is found using the Landauer formula to be $G_{\omega}=4
N_{\omega} e^{2}/ h$. By modulating the parameters of the potential,
one can increase the conductance of the channel from zero in jumps
of $4e^{2}/{h}$. The appearance of the first and further confined
modes within the conducting channel modifies both the strength and
the profile of the potential. This nonlinear screening effect
\cite{Fogler_PRL_100_2008,Fogler_PRB_77_2008} is neglected in the above expression for
$G_{\omega}$ and shall be a subject of future investigation.

Exact solutions for confined modes can also be found for the
P\"{o}schl-Teller potential, $V(x)=-\alpha/\cosh^2(\beta x)$, which
corresponds to $\lambda=2$ in Eq.~(\ref{eq:FinalDiff}).
Wavefunctions can be expressed via Heun polynomials in variable
$\xi=\tanh(\beta x)$.

\section{Discussion and conclusions}
All the results obtained in this paper have been for a specific
potential. However, general conclusions can be drawn from these
results for any symmetric potential. Namely, the product of the
potential strength and its width dictates the number of confined
modes within the channel.\cite{comment}
Moreover, this product has a threshold
value for which the first mode appears. The width of the potential
is defined by the geometry of the top gate structure, and the
strength of the potential is defined by the voltage applied to the
top gate. The mean free path of electrons within graphene is of the
order of 100nm and sub-100nm width gates have been reported in the literature,
\cite{TopGateCombined,Kim_PRL_99_2007,Savchenko_NanoLett_8_2008,Liu_APL_92_2008,GG_PRL_102_2009,Kim_NatPhys_5_2009}
making quantum effects relevant.
The number of modes within such top-gated structures is governed by
the strength of the potential, with new modes appearing with
increasing potential strength. In a top gate structure
 modeled by our potential with a width at half
maximum of 50nm, the first bound mode should appear for a potential
strength of approximately 17\,meV and further modes should appear
with increasing potential strength in steps of 34\,meV, which
corresponds to 395\,K. Therefore a noticeable change in conductivity
should be observed in realistic structures even at room temperature.
This is similar to the quantum Hall effect which is observed in
graphene at room temperature.\cite{QHE} A change of geometry, from
normal transmission to propagation along a potential, allows
graphene to be used as a switching device.

In summary, we show that contrary to the widespread belief, truly
confined (non-leaky) modes are possible in graphene in a smooth
electrostatic potential vanishing at infinity. Full confinement is
possible for zero-energy modes due to the vanishing density of
states at the charge neutrality point. We present exact analytical
solutions for fully confined zero-energy modes in the potential
$V(x)=-\alpha/\cosh(\beta x)$, which provides a good fit (see Appendix)
to experimental potential profiles in existing top gate structures.
\cite{TopGateCombined,Kim_PRL_99_2007,Savchenko_NanoLett_8_2008,Liu_APL_92_2008,GG_PRL_102_2009,Kim_NatPhys_5_2009}
Within such a potential there is a threshold
value of $\omega={\alpha}/{\beta}$ for which bound modes first
appear, which is different to conventional non-relativistic systems.
We found a simple relation between the number of confined modes and
the characteristic potential strength $\omega$.  The threshold
potential strength enables on/off behavior within the graphene
waveguide, and suggests future device applications. The existence of
bound modes within smooth potentials in graphene may provide an
additional argument in favor of the mechanism for minimal
conductivity, where charge puddles lead to a percolation network of
conducting channels.\cite{PercolationNetwork}

There are experimental challenges which need to be resolved in order
to observe confined modes in graphene waveguides. These include
creating narrow gates and thin dielectric layers as well as
optimizing the geometry of the sample to reduce the background
conductance. Our work also poses further theoretical problems,
including the study of non-linear screening, many-body effects,
parity changing backscattering and inter-valley scattering within
the channel.

The study of quasi-one-dimensional channels within conventional
semiconductor systems has lead to many interesting effects. Many
problems are still outstanding, including the 0.7 anomaly in the
ballistic conductance which is a subject of extensive experimental\
and theoretical study.\cite{1DSystem} We envisage that the ability
to produce quasi-one-dimensional channels within graphene will
reveal new and non-trivial physics.

\begin{acknowledgments}
We are grateful to A.V.~Shytov, A.S.~Mayorov, Y.~Kopelevich, J.C.~Inkson,
D.C.~Mattis and N.~Hasselmann for valuable discussions and we thank
ICCMP Bras\'{i}lia and UNICAMP for hospitality. This work was
supported by EPSRC (RRH and NJR), EU projects ROBOCON (FP7-230832) and TerACaN (FP7-230778),
MCT, IBEM and FINEP (Brazil).
\end{acknowledgments}

\appendix*
\section{Potential due to a wire above a graphene sheet}
\begin{figure}[h]
\centering
\includegraphics[width=11cm]{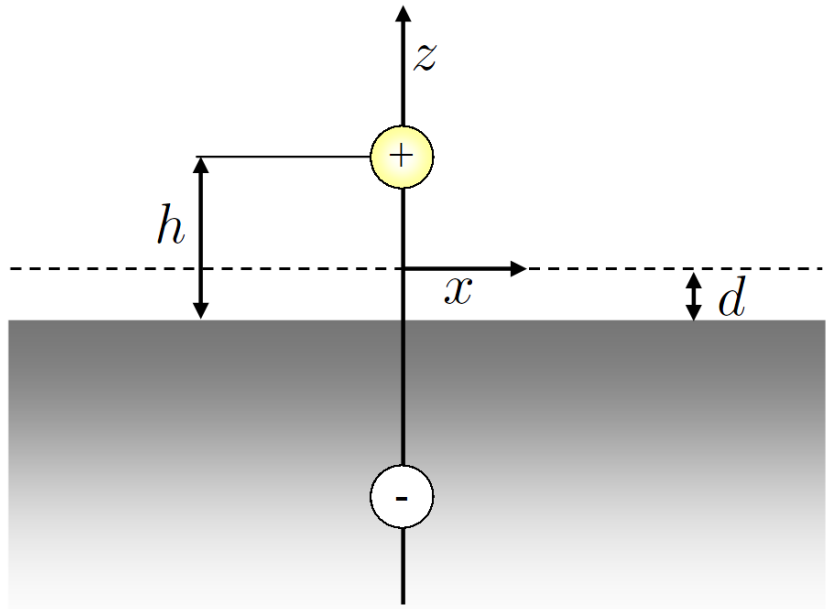}
\caption{The simplified geometry used to obtain the model potential.
The image charge is shown for convenience.} \label{fig:setup}
\end{figure}

To illustrate the relevance of our model potential to realistic
top-gate structures we provide a simple calculation of the potential
distribution in the graphene plane for a simplified top-gate
structure. Fig.~\ref{fig:setup} shows the model that we use for our
estimate. We consider a wire of radius $r_{0}$, separated by
distance $h$ from a metallic substrate (i.e. doped Si) and calculate
the potential profile in the graphene plane separated from the same
substrate by distance $d$. We are interested in the case
when the Fermi level in graphene is at zero energy (the Dirac
point), hence we assume the absence of free carriers in graphene.
We also assume the absence of any dielectric layers, but the problem can be
generalized in such an instance.

\begin{figure}[h]
\centering
\includegraphics[width=9cm, angle=270]{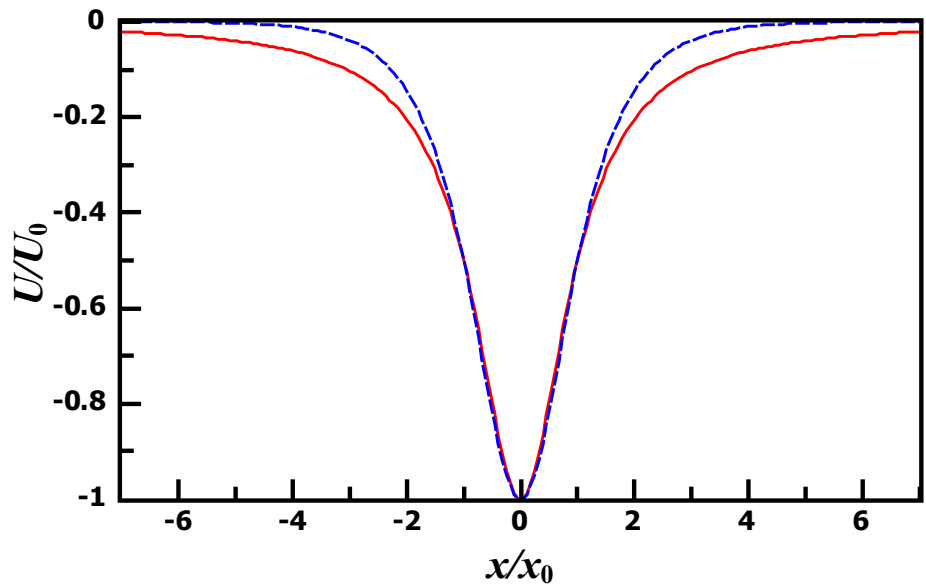}
\caption{A comparison between the potential created by a wire suspended above the graphene plane (solid line) and the $-U_0/\cosh{(\beta x)}$ potential with the same half width at half maximum (dashed line).} \label{fig:potentials}
\end{figure}

One can easily show that the potential energy for an electron in the
graphene plane is given by
\begin{equation}
U(x)=\frac{e\widetilde{\phi_{0}}}{2}\ln\left(\frac{x^{2}+(h-d)^{2}}{x^{2}+(h+d)^{2}}\right),
\label{model1}
\end{equation}
where $\widetilde{\phi_{0}}=\phi_{0}/\ln((2h-r_{0})/r_{0})$,
$\phi_{0}$ is the voltage applied between the top electrode and
metallic substrate and $e$ is the absolute value of the electron charge. One can see that this potential behaves as:
\begin{eqnarray}
  U(x) \approx -U_{0}+{e\widetilde{\phi_{0}}}\;\frac{2hd}{(h^{2}-d^{2})^{2}}\,
  x^{2},
\qquad
  x\ll(h-d);
  \nonumber
  \\
  U(x) \approx -{e\widetilde{\phi_{0}}}\;\frac{2hd}{x^{2}},
  \nonumber
\qquad x\gg(h+d).
\end{eqnarray}
The depth of the potential well is given by
\[
 U_{0}=e\phi_{0}\left[\frac{\ln\left(\frac{h+d}{h-d}\right)}{\ln\left(\frac{2h-r_{0}}{r_{0}}\right)}\right]
 \nonumber
\]
and the half width at half maximum (HWHM) is given by
$x_{0}=\sqrt{h^{2}-d^{2}}$. In Fig.~\ref{fig:potentials} we show a
comparison between the
potential given by Eq.~(\ref{model1}) and the potential considered
in our paper with the same HWHM and potential strength. Clearly the
potential given by $-U_{0}/{\cosh\left(\beta x\right)}$, provides
a significantly better approximation than that of the square well potential.

\end{document}